\documentclass[12pt]{iopart}

\usepackage{graphicx}

\begin{document}

\title{First-principles studies for structural transitions in
ordered phase of cubic approximant Cd$_6$Ca}

\author{K Nozawa \footnote{Present address:
   Graduate School of Material Science,
   University of Hyogo, 3-2-1, Kouto, Kamigori, Hyogo, 678-1297, Japan
  }
 and Y Ishii}

\address
{
Department of Physics, Chuo University, and SORST-JST,
 1-13-27, Kasuga, Bunkyo-ku, Tokyo, 112-8551, Japan
}

\date{\today}

\begin{abstract}
Recently low-temperature structural transition has been reported
for complex cubic compounds Cd$_6$M (M=Ca, Yb, Y, rare earth) and 
it is believed that the transition is due to orientational ordering 
of an atomic shell in the icosahedral cluster in Cd$_6$M.
The first-principles electronic structure calculations and structural
relaxations are carried out to investigate structures and orientational
ordering of the innermost tetrahedral shell of the icosahedral cluster
in Cd$_6$Ca. The very short interatomic distances in the experimental
average structures are relaxed and the innermost tetrahedral shell of 
an almost regular shape is obtained. Three types of orientation for 
the tetrahedral shell and eight different combinations of them for
the clusters at a vertex and body-centre of a cubic cell are obtained.
A possible model describing the orientational ordering at low temperatures
or high pressures is discussed.
\end{abstract}


\maketitle

\section{Introduction}
Cubic Cd$_6$M (M=Yb, Ca) was recognized as an approximant crystal of binary
quasicrystals Cd$_{5.7}$M soon after the discovery of quasicrystalline
phase~\cite{Guo, TsaiNa00, Palenzona, Takakura}.
The crystal structure of Cd$_6$M is understood as a packing of four-layered atomic clusters
with {\it glue} Cd atoms in the interstitial region between them.
Anomalous temperature dependence of the electrical resistivity and specific 
heat was found for cubic Cd$_6$M near 100 K by Tamura {\it et al.}~\cite{TamuraJJAP0}
The anomalies were attributed to orientational ordering
of the innermost shell of the four-layered cluster~\cite{TamuraJJAP, TamuraPRB}.
Watanuki {\it et al.}~\cite{Watanuki} observed various phases under high-pressure, 
where orientation of the innermost shell was assumed to be ordered differently.
The transition is observed so far only for cubic approximants but not for quasicrystals
and cluster linkages in quasiperiodic structures seem to prevent 
the long-range orientational ordering of the innermost shell.
It is certainly important to obtain a microscopic model of the orientational
ordering, not only for elucidating mechanism of the novel structural transition 
in inter-metallic compounds with complex structures
but also for understanding mechanism of quasiperiodic ordering
in Cd-based alloys.

The innermost atomic shell of the four-layered cluster, which is referred to 
as the first shell, is considered to be a tetrahedral Cd cluster~\cite{Palenzona, Gomez}.
The second, third and fourth shells are a dodecahedron of twenty Cd atoms,
an icosahedron of twelve M (=Ca,Yb) atoms and an icosidodecahedron of thirty Cd atoms, respectively.
The four-layered cluster is illustrated in figure~1.
At room temperature, the orientation of the first shell is randomly distributed and
the crystal structure is treated as an average one with the
space group symmetry $Im\bar{3}$.
The four-layered clusters are placed at a vertex and a body-centre of 
a cubic unit cell and the total number of atoms in the unit cell is 
168 including 36 glue atoms.
In X-ray measurements at room temperatures, the first shell 
is described as a fractional site and so 
the structure and orientation of the first shells are open to arguments.

\begin{center}
\begin{figure}[h]
\includegraphics[width=8cm]{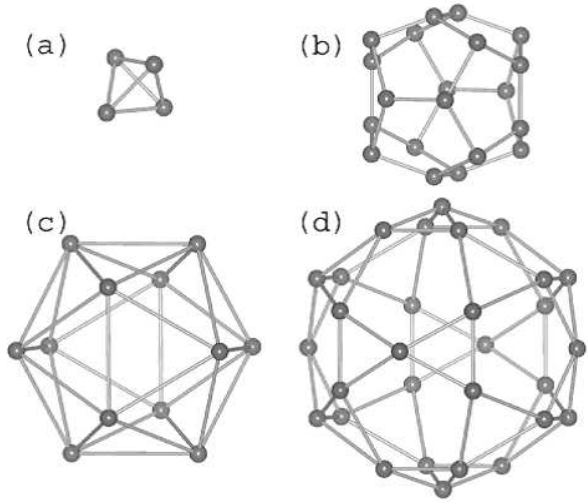}
\caption{
The four-layered atomic cluster of Cd$_6$Ca. The first, second, third and fourth shells
 correspond (a), (b),(c) and (d), respectively.
}
\end{figure}
\end{center}

According to Palenzona's analysis for the high-temperature phase, 
four atoms sit on eight vertices of a small 
cube with half occupancy at the centre of the clusters~\cite{Palenzona}.
Because Cd atoms are not small enough to occupy neighbouring vertices of the cube,
the first shell is reasonably assumed to be of a tetrahedral shape as shown in figure~2(a).
The figures in the right panel of figure~2 is a schematic illustration of the first shells.
Atoms on vertices of a cube means that 
they are under the second shell's atoms on 
the three-fold axis of the cubic unit cell.
Average structures obtained from experiments usually involve
very short interatomic distances. Assuming the atomic structure of Cd$_6$M
proposed by Palenzona as a starting one, we carried out the first-principles
structural relaxation and found that the first and second shells of the
cluster are significantly distorted to avoid the short Cd-Cd distance~\cite{Nozawa}.

\begin{center}
\begin{figure}[h]
\includegraphics[width=7cm]{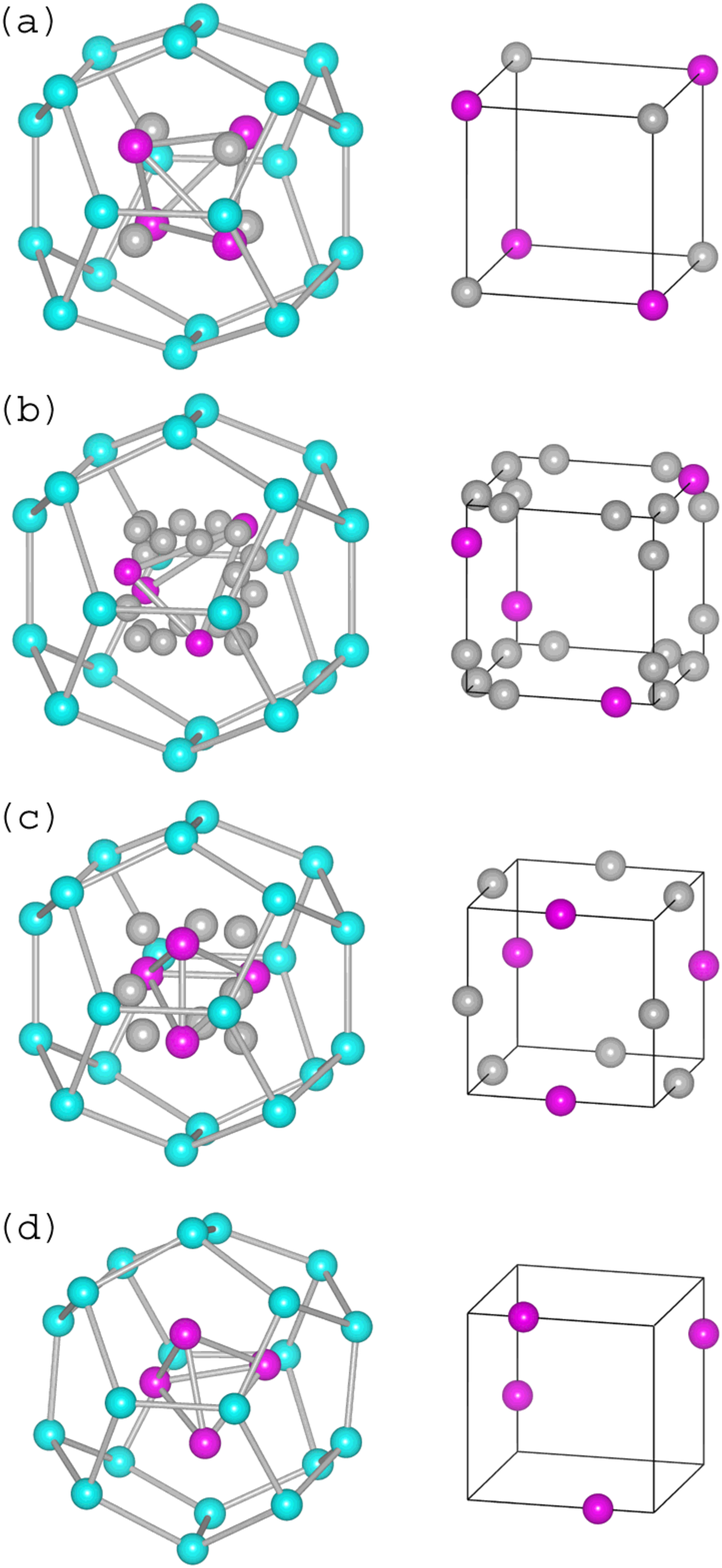}
\caption{
Schematic diagrams for experimentally proposed structures.
(a) In the Palenzona's model, four Cd atoms are sitting on the vertices of a small cube.
 Reasonable choice of occupied vertices forms a regular tetrahedron.
(b) Each vertex of the small cube splits into triple sites and Cd atom occupies the site
with $1/6$ occupancy in the G\'omez-Lidin's model.
(c) Lin-Corbett model is roughly interpreted as the $1/3$ occupancy of the
 midpoint of the edges of the cube. Reasonable choice forms a tetrahedron,
 two-fold axis of which corresponds to that of the outer dodecahedron.
(d) One obtains the Ishimasa type tetrahedron by a rotation of the
 Lin-Corbett type tetrahedron around the two-fold axis.
 Note that the Ishimasa model has no fractional site, because it describes
 the ordered low-temperature phase.
}
\end{figure}
\end{center}

The tetrahedral shape of the first shell leads to a model of orientational ordering,
in which two possible orientations of the tetrahedron 
are treated as Ising spin.
Then the orientational ordering of the tetrahedral first shell is 
a phase transition of the Ising model on a body-centred-cubic lattice.
Ising-like ordering of the tetrahedral shells requires their rotation for orientational changes
and then the energy cost is far larger than thermal energy near the transition 
temperatures reported for ambient- and high-pressure phases~\cite{TamuraJJAP0, Watanuki}.
In fact, we have checked 
that the energy cost is more than 1 eV per cell 
when the tetrahedral first shell at the body centre
is rotated by 90 degrees around the two-fold axis with the first shell 
at the vertex fixed. 
Widom and Mihalkovi\v{c} arrived at a similar conclusion by Ising-model 
analysis for the transition temperature~\cite{Widom}.

G\'omez and Lidin proposed an alternative structural model for a high-temperature
phase of Cd$_6$M, in which each vertex of the tetrahedron is replaced with
triple split sites~\cite{Gomez} (figure~2(b)).
Atoms on the edges in the right panel imply that they are not under the
atoms on the three-fold axis but under the 
pentagonal faces of the second shell.
Since the interatomic distances are longer than those in the 
regular tetrahedron of the Palenzona's structure,
this structural model may be more favourable than the Palenzona's one.
Moreover, the model seems suitable for describing orientational 
changes of the first shell because a rotation of the first shell is not
needed for orientational change.

Similar transition has been found for isostructural Zn$_6$Sc~\cite{Tamura2}.
Structural analyses for the high-temperature disordered phase~\cite{LC} and 
low-temperature ordered one~\cite{Ishimasa} have been reported. 
In the high-temperature phase,
the $24g$ sites ($Im\bar 3$) with $1/3$ occupancy form a cuboctahedron 
as shown in figure~2(c). A reasonable choice of the fractional sites is 
such that gives a tetrahedron as
\begin{equation}
(x, y, 0), \quad (-x, y, 0), \quad (0,-x,-y), \quad (0,-x, y),
\label{coordLC}
\end{equation}
where $x=0.0810$ and $y=0.0748$ by Lin and Corbett~\cite{LC}.
A two-fold axis of the tetrahedron ($y$-axis for the above choice) 
coincides with that of the outer dodecahedral shell and the cubic unit cell. 
We refer this cluster structure as the Lin-Corbett structure hereafter.
In the low-temperature structure, which is for an ordered phase 
with the space group symmetry $C2/2$,
the tetrahedron is rotated slightly around the two-fold axis to
avoid short distances between atoms in the first and second shells.
The atomic positions in this structure are approximately reduced 
to those in a cubic cell as
\begin{equation}
(x, y, z), \quad (-x,  y, -z), \quad (-z, -y, x), \quad ( z, -y,  -x),
\label{coordIS}
\end{equation}
with $x=0.088$, $y=0.064$ and $z=0.018$.
We refer this cluster structure as the Ishimasa structure, which is
illustrated in figure~2(d). 

We reported results of a first-principles structural relaxation of 
cubic Cd$_6$Ca and discussed deformation of the inner atomic shells 
to relax the short interatomic distances in the experimental 
data~\cite{Nozawa, NozawaICQ9}. The previous calculations were, 
however, done with single k-point, and more accurate 
estimates are needed for discussion of the optimal structure. 
Brommer {\it et al.} performed a classical molecular dynamics 
simulation using the potential energies determined by fitting to 
ab-initio data, and found a phase transition near the experimentally 
reported transition temperature~\cite{BrommerAP06}. Although the obtained 
stable cluster is essentially that proposed by Ishimasa {\it et al.}~\cite{Ishimasa}, 
the cluster orientation consistent with the experimental observation was not obtained.
In addition, the previous works were done with a fixed lattice 
constant. The various ordered phase under pressure indicates the favourable 
structure depends on the lattice constant~\cite{Watanuki}. Calculations with different 
lattice constants are therefore required for discussing the phase diagram.

In this paper, we investigate the stable structure of the ordered phase of Cd$_6$M 
at several lattice constants
using the G\'omez-Lidin's structure as a starting
one of the first-principles structural relaxations.

\section{Methods of calculations}

First-principles calculations based on the density functional theory~\cite{DFT} 
 are carried out within the local density approximation~\cite{LDA} to
 determine the stable structure and orientation of the first shells.
The ultra-soft pseudo-potential technique~\cite{UPP}
 are used to represent the effective interaction between the valence electron and ionic core.
The structural (ionic) relaxations are performed as a part of the first-principles
calculations according to the force evaluated as the derivative of the total energy.
Calculations have been performed using the ab-initio total-energy and
molecular-dynamics program VASP (Vienna ab-initio simulation package)
developed at the Institut f\"ur Materialphysik of the Universit\"at
Wien~\cite{VASP1,VASP2, VASP3, VASP4}.

Cd$_6$Yb and Cd$_6$Ca show similar behaviour about the phase transition 
and we study Cd$_6$Ca in this paper.
This is because Ca is easier to be treated in the first-principles
calculation than Yb with localized f-state.
The Cd 4d states are treated as valence states whereas a shallow semicore 
state of Ca 3p is treated as frozen core. Electron-electron interaction
is treated within the local density approximation in the density functional 
theory and the exchange-correlation energy parameterized by Perdew and 
Zunger is used~\cite{PZ81}.
A cubic cell including two four-layered icosahedral atomic clusters and 
36 glue Cd atoms is adopted as a unit cell in all calculations.

The wave functions are expanded with a plane-waves basis set up to a kinetic energy 
cutoff of 168 eV and Kohn-Sham equations are solved iteratively 
to optimize the electronic structure.
The Brillouin zone is sampled with 
14 irreducible k-points (Monkhorst-Pack $3 \times 3 \times 3$ grids). 
The numerical error due to the k-point sampling and the plane-wave cutoff 
is estimated by comparing with the results of more accurate 63 irreducible 
sample k-points (5$\times$5$\times$5 mesh) or 220 eV of plane-wave cutoff. 
The estimated error is of the order of 10 meV for the energy separation of 
different structures at the same lattice constant, while it is 60-80 meV for
different lattice constants.

\section{Results and discussion}

\subsection{Preparations for structural relaxations}
\label{sec1st}
In this section, we explain starting structures for structural relaxations.
According to the analysis by G\'omez and Lidin for the high-temperature
phase of Cd$_6$Ca~\cite{Gomez}, the $48h$ Wyckoff sites for $Im\bar 3$ 
are occupied by Cd with 1/6 probability where coordinates are given as
$$
x=0.08061, \quad y=0.07461, \quad z=0.02687.
$$
This structure is interpreted as that each of the four vertices of the regular 
tetrahedron split into triple sites, which are occupied with 1/3 occupancy.
Therefore the number of possible structures of the first shell is 81 (=$3^4$).
Most of these configurations are, however, equivalent.
The 81 structures are classified into nine inequivalent structures shown 
in figure~3, where the grey and white balls denote the occupied and 
unoccupied sites, respectively.
The number of equivalent structures in each group and six interatomic distances
 in the first shell are listed in table I.

\begin{center}
\begin{figure}[h]
\includegraphics[width=8cm]{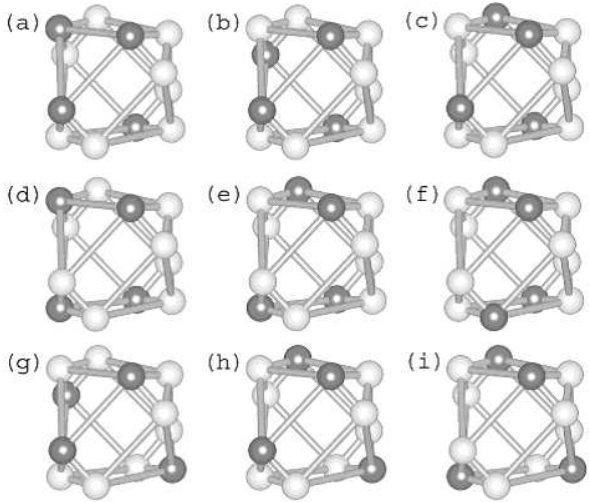}
\caption{
The nine symmetry inequivalent groups of the first shell.
One of the triple split sites is occupied by Cd atoms (grey) and others are vacancies
 (white).
}
\end{figure}
\end{center}

\begin{table}
\begin{center}
\caption{Nine inequivalent structures of the G\'omez-Lidin clusters:
Multiplicity (the number of equivalent configurations), interatomic distances
 and relative total energies are shown.}
{\tabcolsep=1mm
\begin{tabular}{ccccccccc}
Groups & multiplicity &&& Interatomic distances in the first shell [\AA]&&& Energy
 [eV]\\ \hline
\end{tabular}
}
{\tabcolsep=3mm
\begin{tabular}{llrrrrrrrrrr}
 a && 12 && 2.32 & 2.32 & 2.32 & 3.03 & 3.06 & 3.45 & 4.8 \\ 
 b &&  6 && 2.32 & 2.32 & 2.67 & 3.03 & 3.03 & 3.45 & 3.0 \\
 c && 12 && 2.32 & 2.49 & 2.67 & 3.03 & 3.06 & 3.45 & 1.9 \\
 d &&  6 && 2.32 & 2.32 & 2.49 & 3.06 & 3.06 & 3.45 & 3.5 \\
 e && 12 && 2.32 & 2.49 & 2.67 & 3.03 & 3.06 & 3.45 & 1.9 \\
 f &&  3 && 2.49 & 2.49 & 2.67 & 2.67 & 3.45 & 3.45 & 1.1 \\
 g && 12 && 2.32 & 2.67 & 3.03 & 3.03 & 3.03 & 3.06 & 1.1 \\
 h && 12 && 2.32 & 2.49 & 3.03 & 3.06 & 3.06 & 3.06 & 1.6 \\
 i &&  6 && 2.49 & 2.67 & 3.03 & 3.03 & 3.06 & 3.06 & 0.0 \\
\end{tabular}
}
\end{center}
\end{table}

The total energy of each structure is calculated with a cubic unit cell,
in which two identical icosahedral clusters are placed at vertex and body-centre.
Since the structure except for the first shell is identical in each calculation,
 the difference in the total energies can be a measure of relative stability of the first shell.
We find that the group (i) in figure~3 is the most stable structure.
The total energies presented in table I may involve a numerical error of
the order of 10 meV per cell. However, since the difference in the energies 
between the most and second most stable structures is 1.1 eV per cell, 
the numerical error does not influence our conclusion.
We note here that differences in the total energies shown in table I
are originated from differences in interatomic distances between atoms 
in the first shell because those between atoms in the first and 
other shells are identical for all the structures in figure~3.
For the most stable structures group (i), very short Cd-Cd distances
are avoided and six interatomic distances are
close to the nearest neighbour distance, 2.98 \AA, in 
hexagonal-close-packed Cd~\cite{Cd-param}.

The six symmetry-related structures of the group (i) are shown in figure~4
and their atomic coordinates are given as
\begin{eqnarray}
& 1: \quad &
(-x, -y, -z), \quad (-x,  y,  z), \quad ( y,  z, -x), \quad ( y, -z,  x),
\nonumber \\
& 2: \quad &
(-y, -z, -x), \quad (-y,  z,  x), \quad ( x,  y, -z), \quad ( x, -y,  z),
\nonumber \\
& 3: \quad &
(-x, -y, -z), \quad ( x, -y,  z), \quad (-z,  x,  y), \quad ( z,  x, -y),
\nonumber \\
& 4: \quad &
( x,  y, -z), \quad (-x,  y,  z), \quad ( z, -x,  y), \quad (-z, -x, -y),
\label{coordGL}
\\
& 5: \quad &
(-y, -z, -x), \quad ( y,  z, -x), \quad (-z,  x,  y), \quad (-z, -x,  y),
\nonumber \\
& 6: \quad &
( y, -z,  x), \quad (-y,  z,  x), \quad (-z, -x, -y), \quad ( z,  x, -y).
\nonumber
\end{eqnarray}
These structures are equivalent and related with each other by appropriate
symmetry operation. The structure 1 is invariant under two-fold rotation 
around $x$ axis. The structure 2 is derived from 1 by two-fold rotation 
around either $y$ or $z$ axis. The structures 3-6 are obtained by three-fold 
rotation of 1.

\begin{center}
\begin{figure}[h]
\includegraphics[width=8cm]{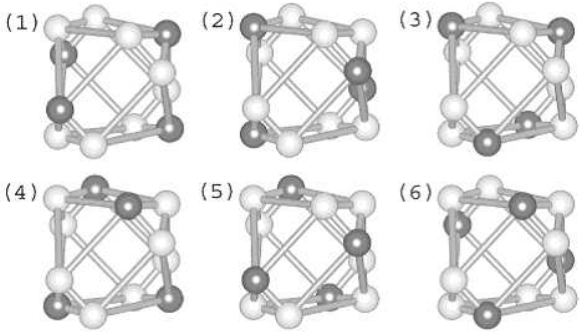}
\caption{
The symmetry equivalent structures of group (i) of figure~3.
White spheres stand for vacancies.
These six structure are isostructural, but face to different directions.
}
\end{figure}
\end{center}

We construct the starting structures using the above stable clusters (i).
Although different unit cells are proposed for Cd-Ca~\cite{Gomez21, Tamuratet} and
Cd-Eu~\cite{GomezEu} systems, we consider here only the cubic unit cell
with two four-layered icosahedral clusters at a vertex and body-centre.
As mentioned above, the most stable structures of the first shell, 
the group (i), has six symmetry-related structures shown in figure~4.
Accordingly, if one fixes the orientation of the first shell at the vertex 
of the cubic unit cell, possible orientation of that at the body-centre is 
one of the six variants in the group (i) and those obtained by
the space-inversion of the group (i).
The 12 configurations are also classified into six inequivalent ones.
They can be denoted using the symbols in figure~4 as
1-1, 1-2, 1-3, 1-1$^*$, 1-2$^*$ and 1-3$^*$.
Here, X$^*$ means the space-inversion of structure X.
For instance, 1-2$^*$ is the combination of structure 1 and space-inversion of
the structure 2.

\subsection{Structural relaxations}
\label{secopt}
After the relaxation,
we find three types of clusters and eight crystal structures
as combinations of the obtained clusters.
Among three types of clusters, two of them are similar to 
the experimentally proposed ones: the Lin-Corbett (LC)~\cite{LC} 
and the Ishimasa (IS)~\cite{Ishimasa} types of structures.
In both of the LC and IS structures, the relaxed positions of 
atoms in the first shell are in the direction of the pentagonal
faces of the dodecahedral second shell to avoid short distances
between atoms in the first and second shells.
Another type of clusters is that never proposed before: One atom is exactly
on the three-fold axis of the outer dodecahedral shell and the cubic
unit cell. A tetrahedron is rotated slightly around the three-fold axis
and the atoms which are not on the three-fold axis are almost in the 
five-fold direction of the dodecahedral shell, $(1,\tau,0)$
where $\tau$ is the golden mean.
Because the first shell's atom on the three-fold axis is capped by 
a pyramid of atoms in the second shell, we refer the new cluster as Mono-capped 
Tetrahedron (MT) in this paper.
Obtained clusters are compared with experimental ones in figure~5.
In figures~5(a) and (b), the grey balls denote obtained cluster and white balls represent
the experimental ones. In the right panel of figure~5(c), a similar
illustration to those in figure~2 is given for the MT type just for comparison.

\begin{center}
\begin{figure}[h]
\includegraphics[width=8cm]{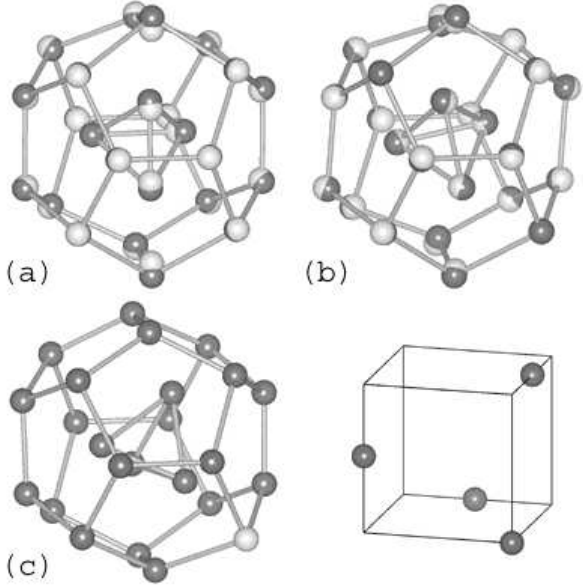}
\caption{
(a) Comparison of the Lin-Corbett structure (white) and calculated 
cluster (grey).
(b) Another type of the relaxed cluster (grey) and the Ishimasa structure (white).
(c) The MT structure, which is different from any previously proposed structures.
 One atom of the first shell is on the three-fold axis of the cubic unit cell,
 whereas the rest three atoms are under the pentagonal faces of the second shell.
 The white ball in the second shell denotes the atom, under which the atom of
 the first shell on the three-fold axis is placed.
}
\end{figure}
\end{center}

The coordinates of the first-shell atoms are summarized in table II.
If one chooses a subset of the fractional sites for describing a 
tetrahedral shell as in (\ref{coordLC}) and (\ref{coordGL}), the tetrahedron
is distorted and a centre of mass of four vertices is shifted from 
the symmetry centre. This is not a case for the Ishimasa structure 
(\ref{coordIS}). In the relaxed structures, distortion of the tetrahedral
shell is removed and a centre of mass of the 
tetrahedral cluster is almost at the symmetry centre.
The interatomic distances in the first shell are
(2.76, 2.83, 2.92$\times$4) for the LC type structures ($a=15.3$ \AA),
(2.83$\times$3, 2.91$\times$3) for the MT type structures ($a=15.3$ \AA) and 
(2.75, 2.80, 2.86$\times$2, 2.89$\times$2) for the IS structures ($a=15.1$ \AA).

\begin{table}
\begin{center}
\caption{Coordinates of atoms in the first shell.}
\begin{tabular*}{70mm}{@{\extracolsep{\fill}}c|ccc}
 & \multicolumn{1}{c}{x} & \multicolumn{1}{c}{y} & \multicolumn{1}{c}{z} \\ \hline
 G\'omez-Lidin~\cite{Gomez} & 0.0806 & 0.0746 & 0.0269 \\ \hline
 Lin-Corbett~\cite{LC} & 0.0810 & 0.0748 & 0 \\ \hline
 Ishimasa~\cite{Ishimasa}       &~0.088 &~0.064 &~0.018 \\
 (translated                   & -0.088 &~0.064 & -0.018 \\
  to the                        & -0.022 & -0.067 &~0.090 \\ 
  cubic lattice)                & ~0.022 & -0.067 & -0.090 \\ \hline
\end{tabular*}
\begin{tabular*}{90mm}{@{\extracolsep{\fill}}c|ccc|ccc}
 & \multicolumn{3}{c|}{vertex}
& \multicolumn{3}{c}{body-centre $+(\frac{1}{2},\frac{1}{2},\frac{1}{2})$} \\
 & \multicolumn{1}{c}{x} & \multicolumn{1}{c}{y} & \multicolumn{1}{c|}{z}
 & \multicolumn{1}{c}{x} & \multicolumn{1}{c}{y} & \multicolumn{1}{c}{z} \\ \hline
      &~0.090 & ~0.075 & ~0.001 & -0.090 & ~0.075 & -0.001 \\
LC(E) &-0.090 & ~0.075 & -0.001 & ~0.090 & ~0.075 & ~0.001 \\
      &~0.000 & -0.066 & -0.093 & ~0.000 & -0.066 & -0.093 \\
      &~0.000 & -0.066 & ~0.093 & ~0.000 & -0.066 & ~0.093 \\ \hline

      & ~0.090 & ~0.073 & ~0.002 & -0.090 & -0.073 & ~0.002 \\ 
LC(I) & -0.090 & ~0.073 & -0.002 & ~0.090 & -0.073 & -0.002 \\
      & ~0.002 & -0.068 & -0.092 & -0.002 & ~0.068 & -0.092 \\
      & -0.002 & -0.068 & ~0.092 & ~0.002 & ~0.068 & ~0.092 \\ \hline

      & ~0.095 & ~0.011 & ~0.059 & -0.095 & -0.011 & ~0.059 \\
MT(2) & -0.060 & ~0.095 & -0.013 & ~0.059 & -0.095 & -0.013 \\
      & ~0.013 & -0.060 & -0.096 & -0.013 & ~0.060 & -0.096 \\
      & -0.071 & -0.070 & ~0.068 & ~0.071 & ~0.070 & ~0.068 \\ \hline

      & ~0.015 & ~0.061 & -0.095 & ~0.061 & ~0.095 & ~0.015 \\
MT(I) & ~0.095 & -0.015 & ~0.061 & -0.095 & ~0.015 & -0.061 \\
      & -0.061 & -0.095 & -0.015 & -0.015 & -0.061 & ~0.095 \\
      & -0.069 & ~0.069 & ~0.069 & ~0.069 & -0.070 & -0.069 \\ \hline

      & ~0.089 & ~0.071 & ~0.019 & -0.089 & -0.071 & ~0.019 \\
IS(2) & -0.089 & ~0.071 & -0.019 & ~0.089 & -0.071 & -0.019 \\
      & -0.018 & -0.068 & ~0.091 & ~0.018 & ~0.068 & ~0.091 \\
      & ~0.018 & -0.068 & -0.091 & -0.018 & ~0.068 & -0.091 \\ \hline

      & ~0.089 & -0.070 & ~0.018 & -0.089 & ~0.070 & -0.018 \\
IS(I) & -0.089 & -0.070 & -0.018 & ~0.089 & ~0.070 & ~0.018 \\
      & ~0.016 & ~0.069 & -0.091 & -0.016 & -0.069 & ~0.091 \\
      & -0.016 & ~0.069 & ~0.091 & ~0.016 & -0.069 & -0.091 \\ \hline

      & -0.017 & -0.066 & ~0.091 & -0.016 & -0.066 & ~0.091 \\
IS(E) & ~0.017 & -0.066 & -0.091 & ~0.017 & -0.066 & -0.091 \\
      & -0.089 & ~0.073 & -0.017 & -0.089 & ~0.073 & -0.017 \\
      & ~0.089 & ~0.073 & ~0.017 & ~0.089 & ~0.073 & ~0.017 \\ \hline

             & ~0.031 & ~0.065 & -0.088 & -0.031 & ~0.065 & -0.088 \\
IS($\sigma$) & -0.031 & ~0.065 & ~0.088 & ~0.031 & ~0.065 & ~0.088 \\
             & ~0.086 & -0.071 & ~0.032 & -0.086 & -0.071 & ~0.032 \\
             & -0.086 & -0.071 & -0.032 & ~0.086 & -0.071 & -0.032 \\ \hline
\end{tabular*}
\end{center}
\end{table}

Depending on the starting structures assumed, we obtain eight inequivalent
crystal structures, in which the clusters at the vertex and body-centre are
of the same type but their orientation is different. Two types of crystal 
structures with the LC-type clusters are obtained. One has 
the LC-type clusters of the same orientation
at the vertex and body-centre of the unit cell and the other has the LC-type 
cluster at the body-centre which is an inversion of that at the vertex.
We refer these structures as LC(E) and LC(I) ones.
If the LC-type cluster rotates around its two-fold axis, the IS-type cluster
is obtained.
For the IS-type clusters, we obtain four 
different structures: (1) the IS clusters at the vertex and body-centre
have the same orientation (IS(E)), (2) the IS cluster at the body-centre 
is obtained by a two-fold rotation of that at the vertex (IS(2))
where the the two-fold rotation is made around an axis perpendicular
to the two-fold symmetry axis of the cluster,
(3) the IS cluster at the body-centre is an inversion of the other (IS(I)),
and (4) the IS cluster at the body-centre is a mirror image of the
other (IS($\sigma$)).
The IS(E) and IS($\sigma$) structures are obtained from the
LC(E) by rotating the first shell around the two-fold axis
whereas the IS(I) and IS(2) ones are obtained from the LC(I).
Therefore the four inequivalent IS structures are classified by combinations of 
the orientation and the rotation direction of the LC cluster.
Finally, two types of orientational combination of the MT clusters are obtained
where the tetrahedra are related with a two-fold rotation in the MT(2) structure 
and an inversion in the MT(I) structure.

We turn our discussion to the stability of the structures.
The structural relaxations are performed at several lattice 
constants to obtain the equilibrium volume.
We have checked accuracy of the total energies for different
lattice constants by changing a cutoff energy.
The optimal lattice constants are around 15.3 \AA\/ for all
the structures and shorter than
the experimental one at room temperatures by about 2.5 \%.
Since a decrease of the lattice constant by thermal expansion is less than
1 \% for temperature difference about 300 K~\cite{TamuraPRB, Tamuratet}, 
this discrepancy is partly because of an error due to the local density approximation.
Although the low-temperature phase of Cd$_6$M is analysed as a
superstructure with a larger unit cell, 
we suppose that the LC, IS and MT clusters are reasonable
candidates for describing the orientational ordering.

The calculated total energies are shown in table III.
The energies are presented as differences from the most stable structure 
(the LC(I) structure at $a=15.3$ \AA).
Note that a symbol '-' indicates that the structure is unstable 
and relaxed to a different one. For instance, IS(2) and IS(I) 
transform to LC(I) structure at 15.3 \AA~.
The relative stability depends not only on structures of the clusters
but on their orientation.
For instance, the LC(I) structure is the most stable one in a wide
range of the lattice constant but the LC(E) structure, which is
 locally identical with the LC(I) structure, takes the highest energy
 among the obtained structures.
This indicates importance of the cluster-cluster interactions.
The most stable structure depends on the lattice constant.
At larger lattice constants than the equilibrium one, 
the MT structure is stable whereas the IS structure becomes more
stable at $a=15.1$ \AA ~or smaller.
It is also interesting to note that the total energies for the MT 
structures with different orientation of the clusters are essentially
the same.

\begin{table}
\begin{center}
\caption{Relative total energies [eV] after the structural relaxations. Asterisks 
represent the optimal energy at each lattice constant. A symbol '-' indicates that the 
structure is unstable.}
\begin{tabular}{cccccccccccccccccc}
& a  && LC(E) && LC(I) && MT(2) && MT(I) && IS(2) && IS(I) && IS(E) && IS($\sigma$)\\ \hline
& 15.7 && 4.854 && 4.819  && 4.724* && 4.727 && ---    && ---  && ---  && --- \\
& 15.5 && 1.522 && 1.470  && 1.435* && 1.440 && ---    && ---  && ---  && --- \\
& 15.4 && 0.536 && 0.471* && 0.476  && 0.479 && ---    && ---  && ---  && --- \\
& 15.3 && 0.082 && 0.000* && 0.047  && 0.052 && ---    && ---  && ---  && --- \\
& 15.1 && 0.819 && 0.700  && ---    && 0.808 && 0.695* && 0.713 && 0.820 && 0.799 \\
& 14.9 && 4.121 && 3.957  && ---    && 4.126 && 3.929* && 3.966 && 4.092 && 4.056 \\
\end{tabular}
\end{center}
\end{table}

In the LC structure, the relaxed positions of 
atoms in the first shell are in the direction of the pentagonal
faces of the dodecahedral second shell.
In figure~6(a), we show the faces of the dodecahedral shell, 
under which the atoms in the tetrahedral first shell is placed,
as coloured ones.
When one chooses four faces as separated as possible from 12 pentagonal 
faces of the dodecahedral second shell, one has faces sharing
an edge inevitably.
A blue (or dark grey in greyscale) face in figure~6(a) is a face sharing an edge with the other
whereas a yellow (or light grey) one is a separated one.
The coordinate of the atom in the first shell under the yellow
face is $(0.002,-0.068,\pm 0.092)$, which is close to the five-fold
direction $(0,-1,\pm\tau)$, whereas the atom under the blue
face is at $(\pm0.090, -0.073, 0.002)$ and shifted from the five-fold 
direction to the green (or grey)
site in the dodecahedral shell. Consequently the green sites 
are moved outward to a position on a triangular face of the
icosahedral third shell. In the IS structure, the distortion
of the second shell is slightly smaller because the rotation 
of the first shell around the two-fold axis relaxes 
the repulsive interaction between atoms in the first
and second shells.

\begin{center}
\begin{figure}[htp]
\includegraphics[width=6cm]{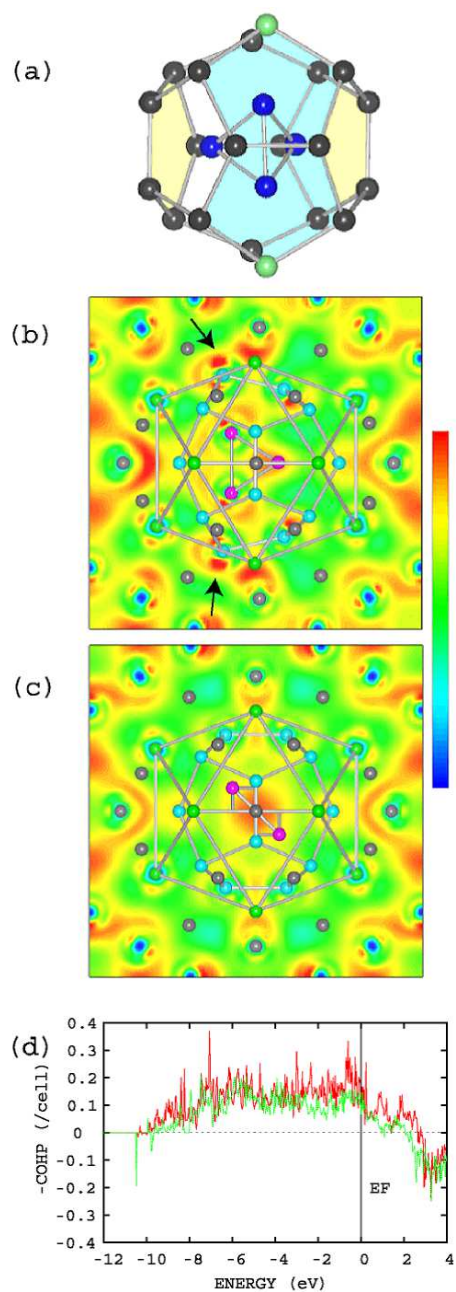}
\caption{
(a) The LC type cluster.
The coloured pentagonal faces indicate those having the first shell 
atoms under themselves.
The blue faces are those sharing an edge and
the yellow ones are separated ones.
The green site represents the atom which is close to the atom in the first 
shell. 
Charge density plots on the (001) plane including (b) the distorted area
of the LC structure and (c) the corresponding area of non-distorted 
second shell with Palenzona's core.
The charge densities are summed over the energy range from -1.0 to -0.5 eV.
Black arrows indicate characteristic charge clouds near the significantly 
distorted site.
(d) The crystal orbital Hamilton population between the displaced 
second shell's atom and the fourth shell's atom locating near the second shell's atom.
The significant increase in the -COHP of the LC structure imply the increase of
Cd-Cd bonding.
}
\end{figure}
\end{center}
 
In figure~6(b), we show the charge densities of the LC(I) structures 
on the (001) plane for the states 
in the energy range from -1.0 to -0.5 eV below the Fermi energy.
The red, blue, green and grey balls in the figure denote the atoms of 
the first, second, third and fourth shells, respectively.
The green sites in figure~6(a) and characteristic charge clouds
around them are indicated by black arrows.
Such charge clouds are not seen around the undistorted second 
shell in the Palenzona's model (figure~6(c)).
Figure 6(d) shows the crystal orbital Hamilton population (COHP)~\cite{COHP}
between the significantly distorted green sites in figure~6(a) and 
the neighbouring atoms in the forth shell.
The calculation is made with the tight-binding linear muffin-tin 
orbitals method in the atomic-sphere approximation~\cite{LMTO}.
The red (solid) curve represents the COHP of the (relaxed) LC(I) structure and
the green (dotted) one is for the unrelaxed structure of the Palenzona's model.
Note that we plot the COHP multiplied -1 in the figure.
Positive values below the Fermi energy imply
the bonding trend between the atoms.
One can find the bonding trend increases for the states
around -1 eV for the relaxed structure.
Therefore the characteristic charge clouds in figure~6(b) 
is a signature of increasing Cd-Cd bonding 
induced by the distortion of the second shell.

The IS type structures, which is obtained by rotating the first shell 
in the LC-type structure around the two-fold axis, become stable at the 
lattice constants smaller than the equilibrium one.
We suppose that the rotation relaxes the repulsive interaction
between atoms in the first and second shells for smaller
lattice constants.
Because the LC(I) structure is more stable than the IS type structure at $a=15.3$ \AA,
it is reasonably assumed that the potential energy surface (PES) with respect to 
the rotation angle of the first shell around the two-fold axis has a single minimum 
as shown in figure~7(a).
In the figure, total energies for unstable structures are evaluated with 
the rotation angle fixed. The solid line is obtained by a cubic spline 
interpolation.
At $a=15.1$ \AA, the IS(2) becomes slightly more stable than
the LC(I) structure where the energy difference is about 5 meV (50 K),
which is as small as a numerical error involved in the present calculation.
Then the PES with respect to the rotation angle have shallow double minima
for $a=15.1$ \AA~ as is shown schematically in figure~7(b). 
The rotation of the first shell is expected to be allowed above 50 K.

\begin{center}
\begin{figure}[h]
\includegraphics[width=8cm]{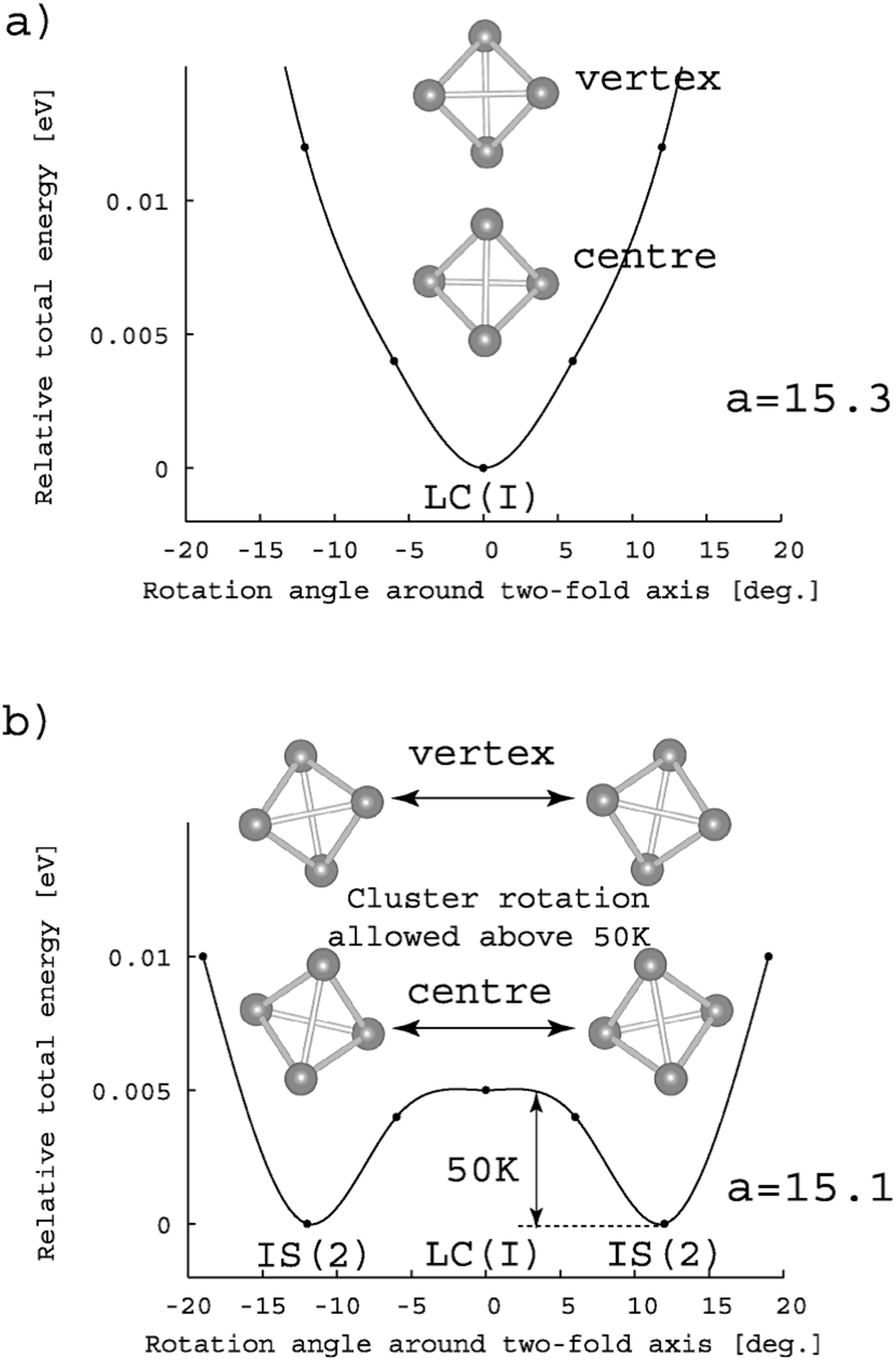}
\caption{
Potential energy surface (PES) with respect to the rotation angle
$\theta$ of the first shell around its two-fold axis.
(a) At the equilibrium lattice constant $a=15.3$ \AA, the LC(I) structure
 is most stable. It implies that the PES has a minimum at $\theta=0$.
(b) At $a=15.1$ \AA ~or smaller, the IS(2) structure becomes the most stable.
Consequently, the PES has two minima at $\theta \ne 0$.
}
\end{figure}
\end{center}

Watanuki {\it et al.} reported various ordered phases in
Cd$_6$Yb at high-pressure~\cite{Watanuki}.
The ordered phase, which is stable at ambient pressure and low temperature (phase I), 
transforms to phase III at 1 GPa.
The compression ratio of the lattice constants at ambient and transition pressure
is about 1 \%~\cite{Watanukipriv}.
By heating, the phase III transforms to phase II at about 130 K via phase III',
which exists in the temperature range about 100-130 K.
We speculate that the IS type structure is stabilized under high pressure
whereas the LC type one is stable at ambient pressure for Cd$_6$Ca.
A ratio of the lattice constants (15.1/15.3) coincides with
the experimental compression ratio of the lattice constant at the transition
from the phase I to III.
Using a volume change by compression $\Delta V$ and $p=1$ GPa, we evaluate
$p\Delta V$ as 0.87 eV/cell, which is comparable with the energy difference,
0.70 eV/cell, between LC(I) at 15.3 \AA~ and IS(2) at 15.1 \AA.
This implies that the transition from the LC structure to the IS one
can take place under the pressure around 1 GPa.

The transition temperature from the phase III to III' (about 100 K) 
is also close to the temperature at which the cluster rotation is allowed (50 K).
Moreover, the energy difference between IS(2) and IS(I) structures at 
$a=15.1$ \AA \/ is about 180 K,
which is close to the transition temperature from the phase III' to II.
Above 180 K, the cluster rotations at the vertex and body-centre 
are expected to take place independently because the thermal energy overcomes 
the energy difference between IS(2) and IS(I).
From these results, we speculate that the phase III is an ordered phase
with the IS-type first shell and the transitions to the phase III' and II
are induced by the cluster rotation of the first shell around the two-fold axis.
Brommer {\it et al.} obtained the IS type cluster as a stable one by 
a classical molecular dynamics simulation, and found a phase transition at 
89 K~\cite{BrommerAP06}. Although the crystal 
structure (orientational configurations of clusters) of the ground state is 
not determined, the reported structure of the stable cluster
is similar to the present result at 15.1 \AA~ and the low transition temperature
is consistent with the proposed PES in Fig.7(b).
The predicted transition by Brommer {\it et al.} therefore might be concerned 
with the cluster rotation around the two-fold axis as shown in figure 7(b).
At $a=14.9$ \AA, the energy difference between the optimal IS(2) and the others
increases. This seems consistent with an increase of the transition temperature
with increasing pressure~\cite{Watanuki}.

In the 1/1 cubic approximant, two types of linkages between the clusters
are realized: One is the two-fold one corresponding to an edge of the 
unit cell and the other is the three-fold one connecting the vertex and
body-centre. These linkages in cubic Cd$_6$Ca are parallel to
the two-fold axis of the LC- and IS-clusters and the
three-fold one of the MT-cluster. 
Takakura {\it et al.}~\cite{Takakuranmat} pointed out significance of
the cluster linkage in  atomic structure of icosahedral Cd-Yb quasicrystal.
Besides the (100) and (111) directions, there are other two- and 
three-fold directions of the linkages in icosahedral quasicrystals.
If the orientation of the tetrahedral shell correlates with 
that of the cluster linkage, the quasiperiodic arrangement
of the clusters may prevent the orientational ordering of the
tetrahedral shell.

\section{Summary and Conclusion}

First-principles structural relaxations are carried out 
for the 1/1 cubic approximant Cd$_6$Ca.
The very short interatomic distances in the experimental 
average structures are relaxed
and the innermost tetrahedral shell of an almost regular shape is obtained.
Three types of orientation of the tetrahedral shell relative to the
second shell are found: the LC-type, IS-type and newly found MT-type.
Although the low-temperature phase of Cd$_6$Ca is analysed as a
superstructure with a larger unit cell, we presume that three types of 
orientation of the tetrahedral shell obtained here are reasonable candidates 
for describing the orientational ordering.

Depending on the starting structures assumed, we obtain eight inequivalent
crystal structures, in which the clusters at a vertex and body-centre are
of the same type but their orientation is different. 
At the equilibrium lattice constant, the LC(I) structure is the most stable.
When the lattice constant decreases,
the IS(2) structure, which is obtained by rotating the first shell 
in the LC(I) structure around the two-fold axis, becomes the most stable one.
It is supposed that the rotation of the first shell relaxes the
repulsive interaction between atoms in the first and second shells
for a smaller lattice constant.
The pressure-induced structural transitions observed by Watanuki {\it et al.}
are discussed in connection with the structural change between the LC and IS 
structures.

The IS-type structure is essentially the same as that obtained by 
Ishimasa {\it et al.} for the low-temperature ordered phase of 
Zn$_6$Sc~\cite{Ishimasa}. In the present calculation for Cd$_6$Ca,
the LC structure is obtained as the most stable one at the
equilibrium lattice constant and the IS one 
becomes the most stable for shorter lattice constants.
As we stressed here, the IS structure is derived from the LC one
by rotating the first shell around its two-fold axis.
We conclude that the LC cluster and its rotation could provide
a plausible model for the orientational ordering in the
complex cubic Cd$_6$M and Zn$_6$Sc.

\section{acknowledgements}

The authors would like to thank K. Makoshi, N. Shima, T. Fujiwara, M. Kraj\v{c}\'{i},
 R. Tamura, T. Watanuki, T. Ishimasa, M. Widom and M. Mihalkovi\v{c} for useful discussion.
Figures of crystal structure and charge density in this paper are drawn by
 VESTA developed by K. Momma and F. Izumi~\cite{VESTA}.
This work is partly supported by Solution Oriented Research for Science and Technology,
 Japan Science and Technology Agency.

\end{document}